\documentstyle{l-aa}     
\def\ref{\par \noindent \hang}

\def\etal{et al.\thinspace}
\def\eg{e.g.\ }

\def\ie{i.e.\ }
\def\approxlt{\mathrel{\hbox{ \lower .5ex \hbox {$\sim$}
	\llap{\raise .15 ex \hbox{$<$}} }}}
\def\approxgt{\mathrel{\hbox{ \lower .5ex \hbox {$\sim$}
	\llap{\raise .15 ex \hbox{$>$}} }}}

\def\multleft#1{\hbox to size{\vbox {\halign {\lft{##}\cr #1}}\hfill}\par}
\def\multright#1{\hbox to size{\vbox {\halign {\rt{##}\cr #1}}\hfill}\par}

\def\degmark{$^\circ$}
\def\today{\ifcase\month\or January\or February\or March\or April\or May\or
      June\or July\or August\or September\or October\or November\or December\fi
      \space\number\day, \number\year}
\def\<{\thinspace}

\def\boxit#1{\vbox{\hrule\hbox{\vrule\kern3pt\vbox{\kern3pt
          #1 \kern3pt}\kern3pt\vrule}\hrule}}

\begin{document}
\begin{sloppypar}

\input psfig.sty
\thesaurus{03(11.01.2; 11.14.1; 11.19.1; 11.19.5; 11.19.6)}

\title{Optical colour maps of Seyfert galaxies. II. More Seyfert 2s\thanks{Based on observations made with the Nordic Optical Telescope, operated on the island of La Palma jointly by Denmark, Finland, Iceland, Norway, and Sweden, in the Spanish Observatorio del Roque de los Muchachos of the Instituto de Astrofisica de Canarias.}}
              
\author{J.K.~Kotilainen\inst{1,2}}

\offprints{J.K.~Kotilainen (Tuorla address)}

\institute{
$^1$ International School for Advanced Studies (SISSA), via Beirut 2--4, 
I--34014 Trieste, Italy\\
$^2$ Tuorla Observatory, University of Turku, V\"{a}is\"{a}l\"{a}ntie 20, 
FIN--21500 Piikki\"{o}, Finland; e-mail: jarkot@deneb.astro.utu.fi
}

\date{Accepted 1998 April 21; Received 1998 April 13; in original form 1998
January 23}

\maketitle

\markboth{J.K ~Kotilainen: Optical colour maps of Seyfert galaxies. II.}{}

\begin{abstract}
We present optical broad band B--I colour maps of a further sample of 10 
Seyfert 2 galaxies. In these bands, the contribution from emission lines to 
the total flux is small, and hence the images predominantly trace the 
continuum distribution. As in our earlier colour maps of a sample of Seyferts 
type 1 and 2, we detect extended blue continuum components in the 
circumnuclear region of several galaxies. These components are either 
elongated (in Mkn 533, Mkn 607, Mkn 1066, NGC 5347, NGC 5953 and NGC 7319) or 
form a double structure across the nucleus (in NGC 5929). They are closely 
aligned with the radio and emission line axes of these galaxies and probably 
arise from scattering of nuclear continuum light by extranuclear mirrors. 
Similar blue elongations (in Mkn 1 and NGC 7212) and double structures (in 
NGC 788) are less well aligned with the radio and line emission, and their 
relationship with scattering regions must be considered uncertain. The 
colours of the blue maxima are consistent with those expected from scattering 
off dust or electrons, a conclusion strengthened by the combined sample of 
the two papers in this series. Our findings thus strongly support the current 
unified models of AGN. 

 
\end{abstract}

\section{Introduction}

In the unified models of active galactic nuclei (AGN), the observed 
differences between various types of AGN are explained in terms of collimated 
radiation, obscuration and viewing angle effects on intrinsically similar 
objects (\eg Antonucci 1993). Specifically for Seyfert galaxies, there is 
increasing evidence (for references see Section 4 and Kotilainen \& Ward 
1997; hereafter KW97), that in Seyfert 2s optically thick material, in the 
form of a thick disk or a dusty torus, blocks our direct view of the compact 
nucleus and the broad line region (BLR), and collimates the ionising photons 
along the minor axis of the torus into two oppositely directed cones.

The emphasis in studies of the nuclear regions of Seyferts has been on the 
emission line properties (\eg integral field spectroscopy and emission line 
imaging). However, although the nuclear torus is expected to be too small to 
be resolved, its outer parts, \eg in the form of a flattened red dust 
distribution, may be detectable in continuum colour maps. Evidence for these 
obscuring regions, perpendicular to the bipolar extended narrow line region 
(ENLR) has been detected in \eg NGC 5252 (Tadhunter \& Tsvetanov 1989; 
Kotilainen \& Prieto 1995) and Mkn 348 (Simpson \etal 1996). Similarly, 
continuum colour maps have revealed blue morphological features coincident 
with the high-excitation gas in the ENLR (\eg Pogge \& De Robertis 1993, 
1995; Mulchaey, Wilson \& Tsvetanov 1996a; KW97). In KW97, we presented high 
resolution colour maps of three Seyfert 1 and six Seyfert 2 galaxies. These 
maps revealed the existence of blue elongations and double structures in NGC 
1068, NGC 3227, NGC 4151, Mkn 3 and Mkn 573, that were not discernible in 
single filter images, and that we interpreted as scattered nuclear light. In 
this paper, we present similar colour maps of the circumnuclear regions of 10 
more Seyfert 2 galaxies. We briefly describe the observations, data reduction 
and the methods used in the construction of the colour maps in section 2. In 
section 3 we present and discuss the morphological features detected for each 
galaxy, including comparison with available multiwavelength data. In section 
4 we discuss the relevance of our findings in the enlarged sample of Seyfert 
galaxies, in view of current unified AGN models. Conclusions are presented in 
section 5. Throughout this paper we use Hubble constant H$_{0}$ = 50 km 
s$^{-1}$ Mpc$^{-1}$, and deceleration parameter q$_{0}$ = 0. 

\section{Observations, data reduction and construction of the colour maps}

The galaxies were observed through broad band B, V, R and I filters, as a 
part of our study of the integrated properties of the host galaxies of 
Seyfert 2 nuclei (Kotilainen 1998). We used the BroCam 1024$^2$ px CCD camera 
with pixel scale 0.176$''$ px$^{-1}$ at the Cassegrain focus of the 2.5 m 
Nordic Optical Telescope (NOT) on La Palma, in August 1995 and July 1996. 
Conditions were photometric throughout with generally subarcsec seeing. The 
images were reduced with standard procedures, \ie bias subtraction, 
flatfielding, residual sky subtraction, and removal of cosmic rays and bad 
pixels. Multiple exposures were coadded after accurate alignment to produce 
the final images. Photometric calibration was obtained against faint standard 
stars selected from Landolt (1992). Airmass correction was applied using 
extinction coefficients appropriate for La Palma. We estimate the internal 
statistical errors from the standard star observations to be $\pm$0.03 mag. 
For further information and for photometry of the galaxies, see Kotilainen 
(1998). 

Colour maps were produced from the B and I band images smoothed to the same 
(lower) resolution, after careful alignment using the centroids of the light 
distribution of field stars as reference points. Misalignments of only a 
small fraction of a pixel may remain, and they do not affect the results for 
structures covering a large number of pixels. The images were smoothed with a 
Gaussian filter with a width corresponding to the seeing, and the I band 
image was divided by the B band image. In the resulting logarithmic colour 
maps, dark shades indicate blue and light shades red emission. At the edge of 
the colour maps noise dominates, but closer to the nucleus the colours are 
reliable. To check that the detected structures in the colour maps are not 
artifacts caused by \eg variable seeing or alignment errors, we have 
inspected the colour residuals of field stars. Reassuringly, they do not show 
any artifacts but cancel out well in the colour maps. Another problem is the 
contamination from emission lines within the broad bands. As discussed in 
KW97, while the V and R filters are severely affected by emission lines, the 
B and I filters are practically devoid of strong emission lines, and are 
dominated by continuum emission. This conclusion is supported by a literature 
search of equivalent widths (EW) of major optical emission lines in the 
nuclear spectra of the sample galaxies. The average contribution of emission 
lines to the B and I filters in this sample are 6 \% and 4 \%, respectively. 
In the next section, we present and discuss for each galaxy a direct B band 
image for morphological information and B--I colour maps (Figs. 1--26). 

\section{Results}

\subsection{Mkn 1} 

Mkn 1 (NGC 449) is a highly inclined SB0a galaxy at redshift z = 0.0159, with 
1$''$ corresponding to 460 pc projected distance in the sky. We show its B 
band image in Fig. 1. The [OIII] emission has radial extent $\sim$9$''$ (4 
kpc) at PA = 83\degmark, elongated along the major axis of the nuclear 
continuum at PA = 87\degmark ~(Mulchaey \& Wilson 1995). Although the 
profiles of optical emission lines of Mkn 1 show blue asymmetry (\eg De 
Robertis \& Shaw 1990), no asymmetry or evident broad wings are seen in the 
near--infrared (NIR; Veilleux, Goodrich \& Hill 1997). There is, therefore, 
no strong evidence for an obscured BLR in Mkn 1. 

We show the B--I map of Mkn 1 in Fig. 2. There are several blue regions in 
the host galaxy to the E and SE of the nucleus, probably related to star 
forming activity, and an arclike red region situated S of the nucleus. Closer 
to the center, there is a blue elongated structure of $\sim$2.5$''$ (1.2 kpc) 
extent across the nucleus at PA = 290\degmark). The nucleus is slightly 
closer to the E end of this elongation. Although not closely parallel to the 
axis of the [OIII] emission (PA = 83\degmark; Mulchaey \& Wilson 1995), this 
blue elongation may be caused by scattered light from the nucleus. We shall 
discuss the origin of the detected structures in all the galaxies, in the 
context of the unified models, in section 4. The colours of the blue 
elongation are given in Table 1. Note that in all the galaxies studied, these 
colours include a contribution from the red underlying stellar population in 
the nuclear region. Therefore, the real colours of the structures must be 
even bluer than those reported here. 

\begin{table*}
\begin{center}
\caption[ ]{The colours of the blue maxima, their distance from the nuclei, and
the relevant position angles.}
\begin{tabular} {llrrrclccc}
\medskip
Galaxy & FWHM & PA(\degmark) & PA(\degmark) & PA (\degmark) & D($''$) & 
D(kpc) & B--V & V--R & V--I\\
\medskip
       & arcsec & ([OIII])     & (radio)      & (B--I)        &         & 
       &     &      &      \\ 
\medskip
(1)      & (2) & (3) & (4)  & (5) & (6) & (7)  & (8)   & (9) & (10)\\
Mkn 1    & 0.8 & 83  &     & 290 & 2.5 & 1.2  & 0.47 & 0.19 & 0.42 \\
Mkn 533  & 0.8 &     & 117 & 310 & 3.2 & 2.7  & 0.99 & 0.32 & 0.75 \\
Mkn 607  & 0.6 & 132 &     & 320 & 0.9 & 0.23 & 1.02 & 0.55 & 1.30 \\
Mkn 1066 & 0.6 & 131 & 134 & 325 & 4.2 & 1.5  & 0.92 & 0.80 & 1.50 \\
NGC 788  & 0.7 & 105 &     & 315 & 1.1 & 0.43 & 1.01 & 0.49 & 1.13 \\
         &     &     &     & 170 & 0.4 & 0.15 & 1.10 & 0.55 & 1.21 \\
NGC 5347 & 0.8 & 0   &     &  10 & 2.6 & 0.60 & 0.69 & 0.51 & 1.18 \\
NGC 5929 & 0.7 & 60  & 61  &  45 & 2.0 & 0.50 & 0.94 & 0.74 & 1.53 \\
         &     &     &     & 225 & 0.8 & 0.20 & 0.99 & 0.86 & 1.38 \\
NGC 5953 & 0.7 & 50  & 40  & 220 & 0.5 & 0.10 & 1.04 & 0.78 & 1.16 \\
NGC 7212 & 0.7 & 10  &     & 165 & 0.4 & 0.31 & 1.35 & 0.23 & 0.68 \\
         &     &     &     & 165 & 1.8 & 1.4  & 1.14 & 0.22 & 0.68 \\
NGC 7319 & 0.8 & 202 & 207 &  10 & 1.1 & 0.72 & 1.29 & 0.74 & 1.46 \\
\end{tabular}
\end{center}
\end{table*}

\subsection{Mkn 533} 

Mkn 533 (NGC 7674, Arp 182) is an SBb spiral galaxy at z = 0.0289, with 1$''$ 
corresponding to 840 pc projected distance in sky. We show the B band image 
of Mkn 533 in Fig. 3. It has asymmetric arms and tidal connection to the 
nearby compact elliptical, NGC 7675. Mkn 533 has a nuclear double radio 
source, with diameter $\sim$0.75$''$ (630 pc) at PA = 117\degmark ~(Kukula 
\etal 1995). H$\alpha$ emission of Mkn 533 has polarised broad wings (Miller 
\& Goodrich 1990; Tran 1995a; Young \etal 1996). The nuclear polarisation 
rises steeply to the blue, the polarised flux spectrum is much bluer than the 
total flux spectrum, and the polarisation PA$\sim$31\degmark ~is independent 
of wavelength, and perpendicular to the radio axis. All this indicates dust 
scattering as the main polarisation mechanism. 

The forbidden optical emission lines (\eg [OIII]) in Mkn 533 have strong blue 
wings (\eg De Robertis \& Shaw 1990; Veilleux 1991) and broad wings have been 
found in the NIR HeI, Pa$\beta$ and Br$\gamma$ lines (Ruiz, Rieke \& Schmidt 
1994; Veilleux \etal 1997). Further evidence for an obscured central source 
in Mkn 533 comes from the hard X-ray spectrum, which shows a prominent Fe K 
line and a flat powerlaw continuum, well fitted by a steep intrinsic spectrum 
reflected by optically thick cold matter (Malaguti \etal 1998). The intrinsic 
X-ray luminosity is at least one order of magnitude larger than what is 
observed. 

We present the B--I maps of Mkn 533 in Fig. 4 (the whole galaxy) and Fig. 5 
(nuclear region). The spiral structure, effects of dust reddening and star 
forming regions are clearly visible in the disk of Mkn 533 (Fig. 4). In the 
nuclear region (Fig. 5), there is a blue elongation from the nucleus to NW 
(PA = 310\degmark) of 3.2$''$ (2.7 kpc) total extent. This structure is 
aligned closely parallel to the radio axis and perpendicular to the 
polarisation orientation, and probably represents scattered light from the 
nucleus. Its colours are given in Table 1. 

\subsection{Mkn 607} 

Mkn 607 (NGC 1320) is a nearly edge--on S0a galaxy at redshift z = 0.0090, 
with 1$''$ corresponding to 260 pc projected distance in the sky. We show a B 
band image of Mkn 607 in Fig. 6. The [OIII] emission is extended along the 
major axis of the host galaxy (PA = 137\degmark) with $\sim$12$''$ (3.1 kpc) 
radius (Mulchaey, Wilson \& Tsvetanov 1996b). The PA of the innermost [OIII] 
emission and that of the nuclear continuum emission is 132\degmark ~(Mulchaey 
\& Wilson 1995; Mulchaey \etal 1996b).

We show the B--I maps of Mkn 607 in Fig. 7 (the whole galaxy) and Fig. 8 
(nuclear region). A red narrow dust lane at closest distance of 5.1$''$ (1.3 
kpc) is visible along the whole SW side of the host galaxy (already seen in 
the B band image; Fig. 6). Several blue star forming regions can be detected 
in the host galaxy (Fig. 7). In the central region (Fig. 8), a red extended 
region is visible on the SW side of the nucleus (at 1.2$''$; 310 pc), 
probably related to the larger scale SW dust lane. In addition, there is a 
faint blue elongated structure emanating from the nucleus toward NW (PA = 
320\degmark) up to 0.9$''$ (230 pc) distance. Its orientation agrees 
perfectly with the axis of the [OIII] emission, and possibly represents 
scattered light from the nucleus. Its colours are given in Table 1. 

\subsection{Mkn 1066} 

Mkn 1066 (UGC 2456) is an inclined SB(s)0 galaxy at redshift z = 0.0121, with 
1$''$ corresponding to 350 pc projected distance on the sky. We show its B 
band image in Fig. 9. The nuclear continuum is aligned at PA = 
137\degmark ~(Mulchaey \etal 1996b). The radial extent of the [OIII] emission 
is 6.5$''$ (2.3 kpc) at PA = 134\degmark, with a double structure across the 
nucleus separated by $\sim$3.3$''$ (1.2 kpc) at PA = 131\degmark ~(Mulchaey 
\etal 1996b). The innermost [OIII] gas is concentrated into a narrow 
jet--like feature extending 1.4$''$ (490 pc) NW of the nucleus at 
PA$\sim310$\degmark, with much fainter emission up to 0.8$''$ (280 pc) SE at 
PA$\sim$130\degmark ~(Bower \etal 1995). The [OIII] emission is thus parallel 
to the axis of the linear triple radio source extending 2.8$''$ (1.0 kpc) 
with bipolar jetlike morphology along PA = 134\degmark ~(Ulvestad \& Wilson 
1989). No polarised or NIR broad lines have been detected in Mkn 1066 (Miller 
\& Goodrich 1990; Kay 1994; Veilleux \etal 1997), and the UV continuum 
polarisation is parallel to the radio axis, unlike in the Seyfert 2s with 
evidence for scattered nuclear light. 

We show the B--I map of Mkn 1066 in Fig. 10. Note that the I band nucleus is 
saturated, so the structure in the central few pixels is not real. The star 
forming regions in the host galaxy are clearly visible in Fig. 10. There is a 
very red region SW of the nucleus at 2.6$''$ (910 pc). Finally, there is a 
blue elongated structure NW of the nucleus (PA = 325\degmark) with total 
extent 4.2$''$ (1.5 kpc). This structure agrees well with the axis of the 
[OIII] and radio emission, and possibly represents scattered light from the 
nucleus. Its colours are given in Table 1. 

\subsection{NGC 788} 

NGC 788 is an S0a galaxy at redshift z = 0.0136, with 1$''$ corresponding to 
390 pc projected distance in the sky. We show its B band image of in Fig. 11. 
It shows faint spiral arms at $\sim$30$''$ (12 kpc) radius from the nucleus. 
The [OIII] emission extends to 4.3$''$ (1.7 kpc) from the nucleus at PA = 
105\degmark, while the major axis of the nuclear continuum is at PA = 
112\degmark ~(Mulchaey \etal 1996b). 

We show B--I maps of NGC 788 in Fig. 12 (the whole galaxy) and Fig. 13 
(nuclear region). The blue inclined ring of emission with $\sim$30$''$ (12 
kpc) radius around the nucleus, tracing the faint spiral structure is clearly 
visible in Fig. 12. The nucleus is situated between two blue regions (Fig. 
13), the less extended at 0.4$''$ (160 pc) S (PA = 170\degmark) and the more 
extended at 1.1$''$ (430 pc) NW (PA = 315\degmark). There is also a red 
extended region at 1.3$''$ (510 pc) S--SW of the nucleus. Although the axis 
of the blue double structure (PA$\sim$150\degmark) does not correspond well 
with the axis of the [OIII] emission (PA = 105\degmark), the blue maxima may 
represent localized peaks in the scattered light from the nucleus. Their 
colours are given in Table 1. 

\subsection{NGC 5347} 

NGC 5347 is an SBab galaxy at redshift z = 0.00778, with 1$''$ corresponding 
to 230 pc projected distance in the sky. Its B band image is presented in 
Fig. 14. It shows a bar enclosed by a ring structure at $\sim$25$''$ (5.8 
kpc) radius, and faint spiral arms emerging from the ring at the ends of the 
bar (see also Gonzalez-Delgado \& Perez 1996). The bar is oriented roughly 
parallel to the major axis of the galaxy. The [OIII] emission of NGC 5347 has 
a double nuclear structure perpendicular to the bar (Pogge 1989). An 
H$\alpha$ emission knot is located at 2.7$''$ (620 pc) NE at PA = 
25\degmark ~from the nucleus, perpendicular to the bar (Gonzalez-Delgado \& 
Perez 1996). The knot has high excitation spectrum and a large Ca triplet EW, 
suggesting the presence of red supergiants associated with an old burst of 
star formation. The emission line ratios of the knot, however, indicate 
photoionisation by a hard AGN continuum (Gonzalez-Delgado \& Perez 1996). The 
nuclear continuum emission is probably emitted anisotropically, as supported 
by photon deficit arguments (Gonzalez-Delgado \& Perez 1996). 

We show the B--I map of NGC 5347 in Fig. 15. The blue star forming regions, 
the ring structure and the spiral arms in the host galaxy are clearly 
visible. There is also a red dust lane running S of the nucleus roughly E to 
W at a closest distance of 3.3$''$ (760 pc), delineating the bar. In the 
central region, the emission knot at 2.7$''$ (620 pc) at PA = 25\degmark ~NE 
of the nucleus is clearly seen. Even closer to the nucleus, there is a blue 
elongated structure with total extent $\sim$1.3$''$ (300 pc) toward N of the 
nucleus (PA = 10\degmark). These structures probably represent the brightest 
localized peaks in the scattered light from the nucleus. Their colours are 
given in Table 1.

\subsection{NGC 5929} 

NGC 5929 is an Sab pec galaxy interacting with the starburst galaxy NGC 5930, 
together forming the galaxy pair Arp 90. The redshift of NGC 5929 is z = 
0.00854, with 1$''$ corresponding to 250 pc projected distance in the sky. 
The B band image of the system is shown in Fig. 16. NGC 5929 has a faint hard 
X--ray spectrum, implying heavy absorption (Rush \& Malkan 1996). It has 
triple radio structure extended along PA = 61\degmark, with total diameter 
1.3$''$ (320 pc; Su \etal 1996). The SW lobe is at 0.7$''$ (170 pc) distance 
from the nucleus and is slightly stronger than the NE lobe at 0.6$''$ (150 
pc) distance. The [OIII] emission resembles closely the radio morphology, 
with two peaks straddling the nucleus at PA$\sim$60\degmark ~(Bower \etal 
1994). However, the [OIII] peaks are closer to the nucleus (1.1$''$ = 280 pc 
separation). Also, the NE [OIII] component is bisected by a dust lane. The 
[OIII] emission does not define a clear biconical morphology as in many other 
Seyfert 2s. Also, there is no evidence for obscuration in the direct images, 
except the dust lane, nor do energy balance considerations suggest anisotropy 
of the nuclear radiation (Bower \etal 1994). 

We show the B--I maps of the whole system in Fig. 17 and of NGC 5929 in Fig. 
18. There is a blue stellar tail from the N part of NGC 5929 toward NGC 5930 
and a red bridge between the galaxies, probably stellar emission reddened by 
the disc of NGC 5930 (Fig. 17; see also Lewis \& Bowen 1993). The nucleus of 
NGC 5929 (Fig. 18) is situated between two blue maxima at opposite sides of 
the nucleus, the closer and brighter at 0.8$''$ (200 pc) SW (PA = 
225\degmark) and the more distant and fainter at 2.0$''$ NE (PA = 
45\degmark). Since the orientation of this structure agrees well with the 
[OIII] and radio axes, and the peaks are only slightly further away from the 
nucleus than the respective [OIII] and radio peaks, they possibly represent 
extranuclear scattering mirrors of anisotropically escaping nuclear light. 
The colours of the blue maxima are given in Table 1. A very red region, 
probably due to dust emission, is located 3.4$''$ (850 pc) S of the nucleus 
(PA = 166\degmark). 

\subsection{NGC 5953} 

NGC 5953 is an S0/a pec spiral galaxy at redshift z = 0.00655, with 1$''$ 
corresponding to 190 pc projected distance in the sky. It interacts with the 
late-type spiral NGC 5954, and together they form the galaxy pair Arp 91 (VV 
244). We show a B band image of the system in Fig. 19. While NGC 5954 has 
pronounced spiral arms, faint emission to N and W of its disk and a stellar 
bridge extending toward NGC 5953, the latter appears relatively smooth. HI 
maps (Chengalur, Salpeter \& Terzian 1995) show a long plume of emission to 
NW of NGC 5953, corresponding to faint optical stellar emission (Fig. 19), 
and probably associated to the S stellar bridge of NGC 5954. 

From long--slit spectroscopy, Yoshida \etal (1993) detected an inclined ring 
of giant star forming regions around the nucleus of NGC 5953, with ring 
diameter $\sim$15$''$ (2.9 kpc). More recently, the circumnuclear star 
forming properties of NGC 5953 were studied by [OIII] and H$\alpha$ imaging 
(Gonzalez-Delgado \& Perez 1997) and by UV imaging (Colina \etal 1997). The 
UV continuum is concentrated in several compact circumnuclear knots in a ring 
with similar morphology to H$\alpha$. Yoshida \etal (1993) detected also a 
high--ionisation region 4$''$ (760 pc) NE from the nucleus at 
PA$\sim50$\degmark. Its location agrees well with radio morphology (PA = 
40\degmark; Jenkins 1984), and it is probably ionised by an anisotropic 
nuclear continuum. Yoshida \etal estimate that the flux seen by the ENLR is 
an order of magnitude higher than that directly observed. The lack of similar 
emission region on the opposite (SW) side of the nucleus is probably due to 
obscuration and/or the torus inclination. 

We show the B--I maps of the whole system in Fig. 20 and of NGC 5953 in Fig. 
21. The overall features already visible in the direct B band image (Fig. 
19), such as the stellar bridge between the galaxies, and the long plume of 
emission to NW of NGC 5953, are even more clearly visible in the B--I map 
(Fig. 20). In NGC 5953, the B--I map (Fig. 21) clearly shows the inclined 
star forming ring oriented roughly N--S with diameter $\sim$13x8$''$ (2.5x1.5 
kpc). There is a red arc around the nucleus E--N--W with two main peaks at 
1.6$''$ (300 pc) at PA = 320\degmark, and at 1.8$''$ (340 pc) at PA = 
60\degmark, inside the star forming ring. Finally, there is a faint narrow 
blue elongated structure of 0.5$''$ (100 pc) extent at PA = 220\degmark. This 
elongation is at similar orientation to the [OIII] and radio axes and 
probably represent scattered light from the nucleus. Its colours are given in 
Table 1. 

\subsection{NGC 7212} 

NGC 7212 is in a pair of interacting galaxies. Its redshift is z = 0.0266, 
with 1$''$ corresponding to $\sim$770 pc projected distance in the sky. We 
show the B band image of NGC 7212 in Fig. 22. Polarised spectra of NGC 7212 
show broad components to H$\alpha$ and H$\beta$, and this polarisation is 
higher than that in the continuum (Tran, Miller \& Kay 1992; Tran 1995a), an 
indication of a hidden BLR. Most likely, the polarisation arises from 
scattering by dust (Tran 1995a), because the spectrum is very red, the 
continuum polarisation rises smoothly to the blue, and the polarised flux 
spectrum is bluer than total flux spectrum. The [OIII] emission of NGC 7212 
shows a jet-like high-ionisation feature extending up to 10$''$ (7.7 kpc) 
from the nucleus at PA = 10\degmark ~(Tran 1995a). This jet is exactly 
parallel to the axis of the small scale double radio source (0.7$''$ 
separation; Falcke, Wilson \& Simpson 1998), and roughly perpendicular to the 
optical polarisation (PA = 93\degmark), suggesting that the jet is collimated 
radiation from the hidden nucleus obscured by a torus. The line ratios of the 
jet indicate photoinisation by the nuclear continuum. There is a faint broad 
base to H$\alpha$, probably the BLR reflected from the obscured nucleus by an 
offnuclear scattering mirror. However, no obvious BLR component to NIR lines 
was found by Veilleux \etal (1997). 

We show the B--I maps of NGC 7212 in Fig. 23 (the whole system) and Fig. 24 
(NGC 7212). The nucleus of NGC 7212 is very blue and an extended fan--shaped 
blue emission region extends from the nucleus to S (PA = 165\degmark) with 
total extent 2.3$''$ (1.8 kpc). This blue region is bisected by a red dust 
lane. A much redder narrow dust lane is situated on the other side of the 
nucleus, at closest distance 3.7$''$ (2.8 kpc) at PA = 280\degmark. Although 
the orientation of the blue elongation does not correspond perfectly with the 
[OIII] emission, it may represent scattered light from the nucleus. Its 
colours are given in Table 1. 

\subsection{NGC 7319} 

NGC 7319 (Arp 319) is an SBb galaxy belonging to the Stephen's Quintet 
compact group of galaxies. Its redshift is z = 0.0225, with 1$''$ 
corresponding to 650 pc projected distance in the sky. We show the B band 
image of NGC 7319 in Fig. 25. In the radio (van der Hulst \& Rots 1981), NGC 
7319 shows a jet-like feature SW to the nucleus at PA = 
207\degmark ~extending $\sim$6$''$ (3.9 kpc). The X--ray emission from 
Stephan's Quintet consists of two components: the soft X-ray emission 
probably arises from hot intracluster gas (Sulentic, Pietsch \& Arp 1995), 
while the hard X-ray emission, peaked on NGC 7319, is consistent with an 
absorbed powerlaw, and exhibits a strong Fe K line, providing strong evidence 
for an obscured nucleus in NGC 7319 (Awaki \etal 1997). 

The [OIII] emission of NGC 7319 extends for $\sim$10$''$ (6.5 kpc) toward 
S--SW, possibly in the form of an ionisation cone (Aoki \etal 1996). The 
[OIII] morphology and PA agree well with the radio, however, the [OIII] peak 
is closer to the nucleus than the radio peak. The kinematics of the extended 
[OIII] emission indicates a high velocity outflow up to $\sim$9$''$ (5.8 kpc) 
from the nucleus, mainly photoionised by the nuclear radiation (Aoki \etal 
1996). The number of ionising photons required to ionise the ENLR is over 10 
times higher than the number of photons in our line of sight, strongly 
indicating an anisotropic nuclear radiation field (Aoki \etal 1996). 

We show the B--I map of NGC 7319 in Fig. 26. To the S of the nucleus (PA = 
210\degmark) there is a very red region at 0.7$''$ (460 pc), continuing 
further S as a more extended red region. The bluest region is situated 
1.1$''$ (720 pc) N of the nucleus (PA = 10\degmark). The orientation of the 
blue elongation agrees well with the [OIII] and radio axes, and probably 
represents scattered light from the nucleus. Its colours are given in Table 1.

\section{Discussion}

Unified models have recently received much emphasis in AGN research (see \eg 
Antonucci 1993). These models postulate that the distinction between broad 
and narrow lined AGN is simply due to our viewing angle. All Seyferts have 
the same basic structure, but in Seyfert 2s the plane of a geometrically and 
optically thick dusty molecular torus lies close to our line of sight and it 
blocks our direct view of the nuclear source and the BLR. On the other hand, 
in Seyfert 1s we look along the axis of the obscuring torus, and directly see 
the nucleus.

Briefly, the evidence in favour of the unified models includes the detection 
in a growing number of Seyfert 2s of: {\bf 1)} polarised broad emission lines 
(\eg Antonucci \& Miller 1985; Miller \& Goodrich 1990; Tran 1995b; Young 
\etal 1996), interpreted as scattered BLR emission by warm material above the 
torus (\eg Krolik \& Begelman 1988; Pier \& Krolik 1992); {\bf 2)} broad NIR 
lines, revealing highly obscured BLRs (\eg Blanco, Ward \& Wright 1990; 
Goodrich, Veilleux \& Hill 1994; Ruiz \etal 1994; Veilleux \etal 1997); {\bf 
3)} the biconical geometry of the high excitation gas (\eg Pogge 1989; 
Tadhunter \& Tsvetanov 1989; Haniff, Ward \& Wilson 1991; Wilson \& Tsvetanov 
1994; Mulchaey \etal 1996a), indicating that the ionizing nuclear radiation 
escapes anisotropically along the torus axis; {\bf 4)} large X--ray absorbing 
column densities and a strong Fe K$\alpha$ emission line (\eg Awaki \etal 
1990; Matt \etal 1996; Malaguti \etal 1998); and {\bf 5)} a deficit of 
directly observed ionizing photons compared to that seen by the ENLR (\eg 
Wilson, Ward \& Haniff 1988; Kinney \etal 1991; Binette, Fosbury \& Parker 
1993). The ENLR (cone) axis, the radio source (torus) axis and the optical 
nuclear continuum axis are usually significantly aligned with each other, 
while there is no correlation with the host galaxy axis at larger scales (\eg 
Pogge \& De Robertis 1993; Wilson \& Tsvetanov 1994; Mulchaey \& Wilson 1995; 
KW97; this paper). 

The colour maps presented in this paper and in KW97 offer a new, independent 
method to test the unified models of AGN. What is the origin of the blue 
elongated or double structures visible in the colour maps of the 
circumnuclear regions of several Seyfert 2 galaxies? In Mkn 3, Mkn 573 and 
NGC 1068 (KW97), and Mkn 533, Mkn 607, Mkn 1066, NGC 5347, NGC 5929, NGC 5953 
and NGC 7319 (this paper), these structures are closely parallel to the radio 
and ENLR emission (to within 20\degmark). In the case of a {\em stellar bar}, 
we would expect {\em red} continuum colours from an old stellar population, 
clearly not seen. {\em Optical synchrotron emission} from the radio jets is 
also unlikely, because there is no detailed correspondence between the blue 
maxima and the radio structure, and the blue continuum is more extended than 
the radio continuum. Alternatively, the blue structures may be due to an {\em 
intrinsically extended nonstellar continuum}, \eg emission from high velocity 
shock waves generated from the interaction of a radio jet with the ENLR gas 
(\eg Sutherland, Bicknell \& Dopita 1993). Such an extended component has 
been proposed to explain the constant H$\beta$ EW over five decades of 
optical--UV continuum luminosity (Binette \etal 1993), and the larger 
polarisation of broad lines than continuum in many Seyfert 2s (Tran 1995b). 
However, a very close morphological correlation between the continuum and the 
high-velocity ionised gas is expected, but not seen in the colour maps. 

Can any of the blue features be due to {\em star formation}? There is now 
increasing evidence for circumnuclear starbursts in many Seyfert 2 galaxies, 
\eg strong far--IR and CO emission from cool dust (Heckman \etal 1989), 
strong extended mid-IR emission and spectral features from warm dust 
(Maiolino \etal 1995), and large NIR light-to-mass ratios (Oliva \etal 1995). 
Also, strong optical CaII triplet 8600 \AA ~absorption with respect to weak 
MgI 5100 \AA ~absorption (Cid Fernandes \& Terlevich 1995), and UV spectral 
properties (Heckman \etal 1995) indicate that hot massive stars in a dusty 
metal-rich starburst can make a significant contribution to the nuclear 
optical/UV energetics of Seyfert 2s. In most Seyferts studied here, the blue 
elongations and double features are closely aligned with the linear radio 
structure and the emission line morphology. Although this could indicate that 
they are intrinsically blue regions of (jet--induced) star formation, we 
consider it unlikely because there is no direct evidence for star forming 
regions in the continuum images, and because the morphology of the regions is 
diffuse, unlike the sharp boundaries and knotty morphology usually seen in 
star forming regions. Furthermore, excitation maps indicate that the 
circumnuclear gas in Seyferts is mainly of high excitation, and does not 
originate from star formation (Mulchaey \etal 1996a). Finally, the detected 
polarised broad lines, heavily absorbed X-ray sources, strong Fe K$\alpha$ 
lines and ionisation cones in an increasing number of Seyfert 2s argue for a 
hidden Seyfert 1 in them.

Therefore, the remaining and most attractive explanation for the blue 
features spatially coincident with the high-excitation circumnuclear gas is 
{\em scattering of the nuclear light} along the radio axis by dust or 
electrons (see also Pogge \& De Robertis 1993; KW97). In either case the 
scattered continuum appears blue, but more so for dust scattering because of 
the wavelength dependence. The reason for the alignment between the continuum 
and the line emission in this case is that both the ionising and optical 
photons escape along the torus polar axis, and the optical continuum is 
scattered by dust and/or electron ``mirrors'' associated with the ionised 
gas. For most of the sample galaxies, the scattering is likely due to 
electrons, because of the shape of the polarised flux spectrum, the 
correlation of the H$\beta$ luminosity with soft X-rays, and the strong Fe 
K$\alpha$ emission and lack of low-energy X-ray absorption (Tran 1995b). 
However, dust is strongly favoured for Mkn 533, NGC 1068 and NGC 7212 (Tran 
1995b; see also KW97). Note that the blue structures in Mkn 1, NGC 788 and 
NGC 7212 are only weakly aligned with the radio and ENLR axes (to within 
20--40\degmark), and their relation with the scattering of the nuclear light 
remains uncertain. 

Finally, in the context of unified models, we might expect to detect red 
regions across the nuclei associated with the dusty torus and perpendicular 
to the radio/ENLR axes (c.f. red nuclear structures in NGC 5252 and Mkn 348; 
Kotilainen \& Prieto 1995; Simpson \etal 1996). Such structures are clearly 
not seen in this sample. This can be understood if the true scale of the 
obscuring material is much smaller than our spatial resolution (as expected 
in the unified model; Mulchaey \etal 1996a). In that case, regions of low 
reddening will be included, and the nuclear reddening will be severely 
underestimated. Thus, while nuclear reddening may be important in Seyferts, 
it remains undetected because of the limited spatial resolution, while the 
larger scale dust lane features become visible, as in many objects of this 
sample. Indeed, if we were to detect reddening by dusty tori in the colour 
maps, these features would correspond to hundreds of pc of scale, much larger 
than expected in unified models. 

\section{Conclusions}

Continuum colour maps provide a powerful method to reveal scattered nuclear 
light in the circumnuclear regions of Seyfert 2 galaxies. As in our previous 
paper (KW97), we have detected elongated or double blue morphological 
structures in the B--I maps of several Seyfert 2 galaxies and found a good 
agreement with previous emission line and radio studies of these galaxies. 
In Mkn 533, Mkn 607, Mkn 1066, NGC 5347, NGC 5929, NGC 5953 and NGC 7319, we 
interpret the colour structure as arising from scattering of the nuclear 
continuum light from electrons or dust in extranuclear mirrors. In Mkn 1, NGC 
788 and NGC 7212, the blue structures are less well in agreement with the 
radio and line emission, and their relationship with scattering regions 
remains uncertain. The colours of the blue maxima are consistent with those 
expected from scattering off dust or electrons, this conclusion strengthened 
by the combined sample of the two papers in this series. Our findings thus 
strongly support the current unified models of AGN. 

\section*{Acknowledgements} 
Thanks go to the referee, M. De Robertis, for several suggestions leading to 
the improvement of the figures, and to R. Falomo for advice in producing the 
final figures. This research has made use of the NASA/IPAC Extragalactic 
Database (NED), which is operated by the Jet Propulsion Laboratory, 
California Institute of Technology.

\section*{References}

\ref Antonucci,R.R.J., 1993, ARA\&A 31, 473
\ref Antonucci,R.R.J., Miller,J.S. 1985, ApJ 297, 621
\ref Aoki,K., Ohtani,H., Yoshida,M., Kosugi,G., 1996, AJ 111, 140
\ref Awaki,H., Koyama,K., Kunieda,H., Tawara,Y., 1990, Nat 346, 544
\ref Awaki,H., Koyama,K., Matsumoto,H., \etal, 1997, PASJ 49, 445
\ref Binette,L., Fosbury,R.A.E., Parker,D., 1993, PASP 105, 1150
\ref Blanco,P.R., Ward,M.J., Wright,G.S., 1990, MNRAS 242, 4P
\ref Bower,G.A., Wilson,A.S., Mulchaey,J.S. \etal, 1994, AJ 107, 1686
\ref Bower,G.A., Wilson,A.S., Morse,J.A. \etal, 1995, ApJ 454, 106
\ref Chengalur,J., Salpeter,E.E., Terzian,Y., 1995, AJ 110, 167
\ref Cid Fernandes,R., Terlevich,R.J., 1995, MNRAS 272, 473 
\ref Colina,L., Garcia Vargas,M.L., Gonzalez-Delgado,R.M. \etal, 1997, ApJ 
488, L71
\ref De Robertis,M.M., Shaw,R.A., 1990, ApJ 348, 421
\ref Falcke,H., Wilson,A.S., Simpson,C., 1998, ApJ, in press
\ref Gonzalez-Delgado,R.M., Perez,E., 1996, MNRAS 280, 53
\ref Gonzalez-Delgado,R.M., Perez,E., 1997, MNRAS 281, 781
\ref Goodrich,R.W., Veilleux,S., Hill,G.J., 1994, ApJ 422, 521
\ref Haniff,C.A., Ward,M.J., Wilson,A.S., 1991, ApJ 368, 167
\ref Heckman,T.M., Blitz,L., Wilson,A.S., Armus,L., Miley,G.K., 1989, ApJ 
342, 735
\ref Heckman,T.M., Krolik,J., Meurer,G. \etal, 1995, ApJ 452, 549 (H95)
\ref Jenkins,C.R., 1984, ApJ 277, 501
\ref Kay,L., 1994, ApJ 430, 196 
\ref Kinney,A.L., Antonucci,R.R.J., Ward,M.J., Wilson,A.S., Whittle,M., 1991, 
ApJ 377, 100
\ref Kotilainen,J.K., 1998, MNRAS, submitted 
\ref Kotilainen,J.K., Prieto,M.A., 1995, A\&A 295, 646
\ref Kotilainen,J.K., Ward,M.J., 1997, A\&AS 121, 77 (KW97)
\ref Krolik,J.H., Begelman,M.C., 1988, ApJ 329, 702
\ref Kukula,M.J., Pedlar,A., Baum,S.A., O'Dea,C.P., 1995, MNRAS 276, 1262
\ref Lewis,J.R., Bowen,D.V., 1993, MNRAS 264, 818
\ref Maiolino,R., Ruiz,M., Rieke,G.H., Keller,L.D., 1995, ApJ 446, 561
\ref Malaguti,G., Palumbo,G.G.C., Cappi,M. \etal, 1998, A\&A 331, 519
\ref Matt,G., Fiore,F., Perola,G.C., \etal, 1996, MNRAS 281, L69
\ref Miller,J.S., Goodrich,R.W., 1990, ApJ 355, 456
\ref Mulchaey,J.S., Wilson,A.S., 1995, ApJ 455, L17 
\ref Mulchaey,J.S., Wilson,A.S., Tsvetanov,Z., 1996a, ApJ 467, 197
\ref Mulchaey,J.S., Wilson,A.S., Tsvetanov,Z., 1996b, ApJS 102, 309
\ref Oliva,E., Origlia,L., Kotilainen,J.K., Moorwood,A.F.M., 1995, A\&A 301, 
55 
\ref Pier,E.A., Krolik,J.H., 1992, ApJ 401, 99
\ref Pogge,R.W., 1989, ApJ 345, 730
\ref Pogge,R.W., De Robertis,M.M., 1993, ApJ 404, 563
\ref Pogge,R.W., De Robertis,M.M., 1995, ApJ 451, 585
\ref Ruiz,M., Rieke,G.H., Schmidt,G.D., 1994, ApJ 423, 608
\ref Rush,B., Malkan,M.A., 1996, ApJ 456, 466
\ref Simpson,C., Mulchaey,J.S., Wilson,A.S., Ward,M.J., Alonso-Herrero,A., 
1996, ApJ 457, L19
\ref Su,B.M., Muxlow,T.W.B., Pedlar,A. \etal, 1996, MNRAS 279, 1111
\ref Sulentic,J.W., Pietsch,W., Arp,H., 1995, A\&A 298, 420
\ref Sutherland,R.S., Bicknell,G.V., Dopita,M.A., 1993, ApJ 414, 510
\ref Tadhunter,C., Tsvetanov,Z., 1989, Nat 341, 422
\ref Tran,H.D., 1995a, ApJ 440, 578
\ref Tran,H.D., 1995b, ApJ 440, 597
\ref Tran,H.D., Miller,J.S., Kay,L.E., 1992, ApJ 397, 452
\ref Ulvestad,J.S., Wilson,A.S., 1989, ApJ 343, 659
\ref van der Hulst,J.M., Rots,A.H., 1981, AJ 86, 1775
\ref Veilleux,S., 1991, ApJ 369, 331
\ref Veilleux,S., Goodrich,R.W., Hill,G.J., 1997, ApJ 477, 631
\ref Wilson,A.S., Tsvetanov,Z., 1994, AJ 107, 1227
\ref Wilson,A.S., Ward,M.J., Haniff,C.A., 1988, ApJ 334, 121
\ref Young,S., Hough,J.H., Efstathiou,A., \etal, 1996, MNRAS 281, 1206
\ref Yoshida,M., Yamada,T., Kosugi,G., Taniguchi,Y., Mouri,H., 1993, PASJ 45, 
761

\section*{Figure Captions}

\noindent{\bf Fig. 1.} The B band image of Mkn 1. The size of the image is 
26x26$''$ (12x12 kpc). In this and other Figures, north is up, east to the 
left, the nucleus of the Seyfert galaxy is marked as a cross and, unless 
otherwise indicated, the scale bar corresponds to 1 kpc projected distance in 
the sky.

\noindent{\bf Fig. 2.} The B--I map of Mkn 1. The size of the image is 
26x26$''$ (12x12 kpc). In this and other colour maps, dark shades indicate 
blue and light shades red emission. The B--I colour coding is from 1.0 to 
2.0. Note the blue elongation across the nucleus at PA = 290\degmark ~of 
$\sim$2.5$''$ (1.2 kpc) extent.

\noindent{\bf Fig. 3.} The B band image of Mkn 533. The size of the image is 
79x79$''$ (67x67 kpc). The scale bar corresponds to 5 kpc projected distance 
in the sky. 

\noindent{\bf Fig. 4.} The B--I map of Mkn 533. The size of the image is 
79x79$''$ (67x67 kpc). The B--I colour coding is from 1.2 to 2.4. The scale 
bar corresponds to 5 kpc projected distance in the sky. Spiral structure and 
star forming regions are clearly visible in the main body of the galaxy. 
There is a blue elongation with a total extent of 3.2$''$ (2.7 kpc) from the 
nucleus along PA = 310\degmark.

\noindent{\bf Fig. 5.} The B--I map of the innermost 21x21$''$ (18x18 kpc) of 
Mkn 533, showing more clearly the structures seen in Fig. 4. The B--I colour 
coding is from 0.7 to 2.4.

\noindent{\bf Fig. 6.} The B band image of Mkn 607. The size of the image is 
53x53$''$ (14x14 kpc).

\noindent{\bf Fig. 7.} The B--I map of Mkn 607. The size of the image is 
53x53$''$ (14x14 kpc). The B--I colour coding is from 2.2 to 2.8. Note the 
red dust lane along the whole SW side of the galaxy. Star forming regions are 
visible in the main body of the galaxy.

\noindent{\bf Fig. 8.} The B--I map of the innermost 18x18$''$ (4.7x4.7 kpc) 
of Mkn 607. The B--I colour coding is from 2.2 to 2.8. Note the possible blue 
elongation emanating at PA = 320\degmark ~up to 0.9$''$ (230 pc) distance from 
the nucleus. There is a red region on the SW side of the nucleus, probably 
related to the larger scale dust lane (Fig. 7).

\noindent{\bf Fig. 9.} The B band image of Mkn 1066. The size of the image 
is 53x53$''$ (19x19 kpc).

\noindent{\bf Fig. 10.} The B--I map of Mkn 1066. The size of the image is 
53x53$''$ (19x19 kpc). The B--I colour coding is from 1.6 to 3.0. Note that 
the I band image is saturated in the nucleus, so the structure in the central 
few pixels is not real. There is a blue elongated region at PA = 
326\degmark ~from the nucleus with 2.6$''$ (910 pc) extent, and a very red 
region SW of the nucleus. Spiral arms and star forming regions are visible in 
the disk.

\noindent{\bf Fig. 11.} The B band image of NGC 788. The size of the image 
is 70x70$''$ (27x27 kpc).

\noindent{\bf Fig. 12.} The B--I map of NGC 788. The size of the image is 
70x70$''$ (27x27 kpc). The B--I colour coding is from 2.2 to 2.6. Note the 
blue inclined ring of emission at $\sim$30$''$ (12 kpc) radius from the 
nucleus (see also Fig. 11). The nucleus is situated between two blue regions, 
at 0.4$''$ (150 pc) along PA = 170\degmark ~and at 1.1$''$ (430 pc) along PA = 
316\degmark. There is also an extended red region SW of the nucleus.

\noindent{\bf Fig. 13.} The B--I map of the innermost 8.8x8.8$''$ (3.4x3.4 
kpc) of NGC 788, showing more clearly the structures seen in Fig. 12. The 
B--I colour coding is from 2.2 to 2.6.

\noindent{\bf Fig. 14.} The B band image of NGC 5347. The size of the image 
is 88x88$''$ (20x20 kpc).

\noindent{\bf Fig. 15.} The B--I map of NGC 5347. The size of the image is 
88x88$''$ (20x20 kpc). The B--I colour coding is from 0.7 to 2.1. Note the 
red dust lane running S of the nucleus roughly from E to W, probably related 
to the bar, and a blue elongation of $\sim$2.6$''$ (600 pc) extent N of the 
nucleus at PA = 8\degmark.

\noindent{\bf Fig. 16.} The B band image of NGC 5929/5930. The size of the 
image is 70x70$''$ (18x18 kpc). NGC 5929 is situated in the SW and NGC 5930 
in the NE part of the Figure.

\noindent{\bf Fig. 17.} The B--I map of NGC 5929/5930. The size of the image 
is 70x70$''$ (18x18 kpc). The B--I colour coding is from 1.0 to 2.4. NGC 5929 
is situated in the SW and NGC 5930 in the NE part of the Figure. The blue 
stellar tail from the N part of NGC 5929 toward NGC 5930 and the red stellar 
bridge between the galaxies are clearly visible. The nuclear region of NGC 
5929 has double blue peaks at opposite sides of the nucleus, at 0.8$''$ (200 
pc) at PA = 227\degmark, and at 2.0$''$ (500 pc) at PA = 47\degmark). There 
is also a very red region S of the nucleus.

\noindent{\bf Fig. 18.} The B--I map of the innermost 18x18$''$ (4.5x4.5 kpc) 
of NGC 5929, showing more clearly the structures seen in Fig. 17. The B--I 
colour coding is from 1.0 to 2.3.

\noindent{\bf Fig. 19.} The B band image of NGC 5953/5954. The size of the 
image is 106x106$''$ (20x20 kpc). NGC 5953 is situated in the SW and NGC 5954 
in the NE part of the Figure.

\noindent{\bf Fig. 20.} The B--I map of NGC 5953/5954. The size of the image 
is 106x106$''$ (20x20 kpc). The B--I colour coding is from 0.2 to 2.2. NGC 
5953 is situated in the SW and NGC 5954 in the NE part of the Figure. Note 
the blue stellar bridge between the galaxies, and the long plume of blue 
emission NW of NGC 5953. There is a red arc around the nucleus of NGC 5953 
E--N--W, and an inclined lopsided star forming ring roughly N--S with 
diameter 13x8$''$ (2.5 x 1.5 kpc). Finally, there is possibly a blue 
elongated structure up to 0.5$''$ (100 pc) from the nucleus at PA = 
221\degmark.

\noindent{\bf Fig. 21.} The B--I map of the innermost 18x18$''$ (3.4x3.4 kpc) 
of NGC 5953, showing more clearly the structures seen in Fig. 20. The B--I 
colour coding is from 0.7 to 2.2.

\noindent{\bf Fig. 22.} The B band image of the field of NGC 7212. The size 
of the image is 44x44$''$ (34x34 kpc). NGC 7212 is situated in the SW and the 
main companion galaxy in the NE part of the Figure. The scale bar corresponds
to 5 kpc projected distance in the sky. 

\noindent{\bf Fig. 23.} The B--I map of the field of NGC 7212. The size of 
the image is 44x44$''$ (34x34 kpc). The B--I colour coding is from 1.4 to 
3.1. NGC 7212 is situated in the SW and the main companion galaxy in the NE 
part of the Figure. The scale bar corresponds to 5 kpc projected distance in 
the sky. The nuclear region of NGC 7212 is very blue and an extended 
fan--shaped blue region emanates from the nucleus along PA = 166\degmark, 
with total extent of 2.3$''$ (1.8 kpc). This blue region is bisected by a red 
dust lane. A red narrow dust lane is visible on the other side of the nucleus.

\noindent{\bf Fig. 24.} The B--I map of the innermost 18x18$''$ (14x14 kpc) 
of NGC 7212, showing more clearly the structures seen in Fig. 23. The B--I 
colour coding is from 1.4 to 3.1. The scale bar corresponds to 5 kpc projected
distance in the sky. 
 
\noindent{\bf Fig. 25.} The B band image of NGC 7319. The size of the image 
is 35x35$''$ (23x23 kpc). The scale bar corresponds to 5 kpv projected 
distance in the sky. 
 
\noindent{\bf Fig. 26.} The B--I map of NGC 7319. The size of the image is 
35x35$''$ (23x23 kpc). The B--I colour coding is from 2.4 to 3.0. The scale 
bar corresponds to 5 kpv projected distance in the sky. The bluest region 
lies at 1.1$''$ (720 pc) from the nucleus at PA = 10\degmark. To the S of the 
nucleus at 0.7$''$ (460 pc) there is a very red region, extending further S 
as a red region.

\end{sloppypar}
\end{document}